\documentclass[twocolumn,tighten,time]{aastex631}
\usepackage[fleqn]{amsmath} 
\usepackage{graphicx}	
\pdfoutput=1 
\usepackage{epsfig}
\usepackage{subfigure}
\usepackage{multirow}

\newcommand{\der}{{\rm d}}

 \newcommand{\cc}{_{\rm c}}
\newcommand{\co}{^{\rm co}}

 \newcommand{\h}{_{\rm h}}

\newcommand{\F}{^{\rm th}}

\newcommand{\rs}{r_{\rm c}} \newcommand{\rhos}{\rho_{\rm c}}
\newcommand{\maxi}{_{\rm max}} 
 \newcommand{\modotc}{M$_\odot$}
\newcommand{\ti}{t_{\rm i}} \newcommand{\e}{_{\rm e}}

\newcommand{\beq}{\begin{equation}} \newcommand{\eeq}{\end{equation}}
 \newcommand{\m}{_{\rm m}}
 \newcommand{\beqa}{\begin{eqnarray}}
\newcommand{\eeqa}{\end{eqnarray}} \newcommand{\lav}{\langle}
\newcommand{\rav}{\rangle} 
\newcommand{\vir}{_{\rm vir}}

\begin{document}

\shorttitle{Origin and Characterization of the Secondary Halo Bias}
\shortauthors{Salvador-Sol\'e, Manrique, \& Agulló}

\title{ORIGIN AND FULL CHARACTERIZATION OF THE SECONDARY (ASSEMBLY) HALO BIAS}

\author{Eduard Salvador-Sol\'e, Alberto Manrique, and Eduard Agulló}
\affiliation{Institut de Ci\`encies del Cosmos. Universitat de Barcelona, E-08028 Barcelona, Spain}

\email{e.salvador@ub.edu}



\begin{abstract}
The clustering of dark matter halos depends not only on their mass, the so-called primary bias, but also on their internal properties, the so-called secondary bias. While the former effect is well-understood within the Press-Schechter (PS) and excursion set (ES) models of structure formation, the latter is not. In those models, protohalos are fully characterised by their height and scale, which determine the halo mass and collapse time, so there is no room for any other halo property. This is why the secondary bias was believed not to be innate but due to the distinct merger rate of halos lying in different backgrounds, and dubbed assembly bias. However, it is now admitted that mergers leave no imprint in the inner halo properties. In fact, the innate origin of the secondary bias cannot be discarded because, in the more realistic peak model of structure formation, halo seeds are characterized by one additional property: the peak curvature. Here we use the confluent system of peak trajectories (CUSP) formalism to show that peaks lying in different backgrounds have different mean curvatures, which in turn cause them to evolve in halos with different typical inner properties. The dependence we find of the properties on halo background (or halo clustering) reproduces the results of simulations. 
\end{abstract}

\keywords{methods: analytic --- cosmology: theory, dark matter --- dark matter: halos --- galaxies: halos}


\section{INTRODUCTION}\label{intro}

One fundamental property of the Universe is that light does not trace matter. More massive galaxies are more clustered than less massive ones \citep{Cea80} and galaxies of a given mass but different mass-to-light ratios, morphologies or star formation rates are also differently clustered \citep{FG79,D80,Pea99}. As the spatial distribution of cosmic objects informs on their formation and evolution, the discovery of those segregations gave rise to a lively ``nature vs. nurture'' debate \citep{SS90,Dea97,Eea97,Sea05}. In fact, galaxies form and develop within dark matter (DM) halos, which are themselves segregated (biased) in mass as well as in other internal properties, so the spatial distribution of galaxies is tightly coupled to that of halos and their substructure (e.g. \citealt{PS00,S00,Zea05,Mea20,XZ20}). 

It is now well-established that the mass segregation of halos \citep{HP73,BS83}, the so-called primary bias, is already imprinted in the primordial density field (\citealt{K84,BBKS}, hereafter BBKS). However, the reason why halos with the same mass but different internal properties are also differently clustered, the so-called secondary bias, is poorly understood.

The first evidence in simulations of the secondary bias was found by \citet{ST04}, who noticed that halos in pairs form earlier than more isolated ones.  \citet{Hea06} showed that, in general, the more strongly clustered halos, the smaller their formation time, an effect which was confirmed by other authors \citep{Gea05,Hea06,Ze06,Wech06,Wet07,Jea07}. In addition, \citet{Wech06} found that low-mass halos, with the highest concentration, are more clustered, while the opposite is true for high-mass halos. Lastly, \citet{GW07} showed that the secondary bias affects other internal properties of haloes such as their peak velocity, subhalo abundance (or halo occupation number) and spin. Those trends were subsequently confirmed and extended to other properties such as the velocity anisotropy and shape (\citealt{Mea07,A08,FW10,Lea17,Mea18,SP19,Cea20,Lea23}; see also \citealt{RP20,Mea20,Hea21}). 

In the PS \citep{PS} and excursion set (ES; \citealt{BCEK}) models of structure formation, using top-hat and $k$-sharp smoothing windows, respectively, protohalos are fully characterized by their height and scale, which determine the mass and collapse time of the associated halos regardless of their environment. Thus there is no room in such models for halos of a given mass at a given time to have distinct internal properties according to their background \citep{GW07}. The fact that the merger history of halos depends on their environment \citep{Gea01,Gea02,FM09,FM10,Wet07} suggested that the secondary bias was the result of the distinct assembly (merging) history of halos with identical seeds but evolving in different environments. This is why it was dubbed ``assembly bias''. However, neither the different merger rate of halos evolving in different environments nor any other evolutionary mechanism explored \citep{Mea05,Sea07,D08,Hea09,Yea17} could reproduce the observed properties of the secondary bias.

Furthermore, numerical experiments (\citealt{Mea99,Hea99,Haea06,WW09,Bea12}; but see \citealt{HT10,Wea20a}) showed that mergers leave no imprint in halo properties (except in their substructure). The result found by \citet{WW09} is particularly compelling in the context of the assembly bias: {\it all internal properties of halos except subhalo abundance are identical in purely accreting halos} (i.e. having undergone monolithic collapse) {\it than in halos grown hierarchically} (i.e. having suffered major mergers). In addition, \citet{Mea18} and \citet{Cea20} found that the secondary bias does not correlate with the assembly (or merger) history of halos. Last but not least, \citet{SM19} formally proved such a fundamental characteristic of structure formation. 

\citet{Z07} showed that the ES model with a smoothing filter different than the $k$-sharp one so that protohalos are sensitive to the matter distribution on larger scales leads to halos with formation times dependent on the background. Although, that generic result did not explain the secondary bias found in simulations, it showed that an innate origin of the secondary bias was possible provided the suited model of structure formation. \citet{Dea08} noted that, in the peak model where halos form from the collapse of patches around density maxima (peaks) in the smoothed linear Gaussian random density field, protohalos are characterised not only by their height at a given scale, but also by their curvature (i.e. minus the Laplacian of the density field at the peak scaled to its rms value),\footnote{Peaks also have different ellipticities and prolatenesses, but the typical values of these properties depend on their typical curvature (BBKS).} which depends on the peak background. Consequently, halos could have different internal properties and formation times arising in a simple natural manner from the specific properties of their seeds, regardless of their assembly history or any other environmental effect. 

Unfortunately, some difficulties met in the peak model (see \citealt{PI}, hereafter Paper I) have so far prevented from checking this. If the derivation of the primary bias in the peak model was already challenging, examining whether the secondary bias can be explained in that theoretical framework was even harder: it requires, in addition, connecting the curvature of peaks with the typical internal properties of halos. But that is now feasible thanks to the {\it ConflUent System of Peak trajectories} (CUSP) formalism, that establishes such a connection from first principles and with no free parameter \citep{SM19}. 

In Paper I we applied CUSP to derive in a robust way the primary bias in the peak model. Here we use it to explain and characterize the secondary bias. The layout of the Paper is as follows. In Section \ref{correspondence} we remind the basics of the CUSP formalism. In Section \ref{density}, we compare the average density profile of unconstrained halos to that of halos lying in a background. The consequences of this relation on the different internal properties of halos involved in the secondary bias are examined in Section \ref{assembly}. Our results are summarised and the main conclusions are drawn in Section \ref{dis}.

\section{Halo-Peak Correspondence}
\label{correspondence}

CUSP allows one to derive all macroscopic halo properties from peak statistics in the linear random Gaussian density field. This is possible thanks to the fact that there is a one-to-one correspondence between halos with $M$ at $t$ (for any particular mass definition) and peaks with density contrast $\delta$ in the linear density field at the initial (arbitrary) time $\ti$, smoothed with a Gaussian filter of scale $R$. The reader is referred to \citet{SM19} for details. Next we provide a quick overview of this correspondence.

The time $t$ of ellipsoidal collapse of patches around peaks depends not only on their mass and size as in top-hat spherical collapse, but also on their triaxial shape and concentration \citep{P69}. Nonetheless, the probability distribution functions (PDFs) of ellipticities $e$, prolatenesses $p$, and curvatures $x$ of peaks with $\delta$ at scale $R$ are very sharply peaked (BBKS), so peaks with any given $\delta$ at $R$ have essentially the same fixed values of those quantities, implying that they collapse (and virialize) essentially at the same time $t$, dependent only on $\delta$ at a fixed scale $R$. Consequently, given any mass definition of halos fixing their mass $M$ at $t$, the scale $R$ of the corresponding peaks with $\delta(t)$ at $\ti$ define the relation between $R$ and $M$, dependent on $t$ in general, for peaks collapsing in halos with $M$ at $t$. 

As shown by \cite{Jea14a}, the relations $\delta(t)$ and $R(M,t)$ can be found, for any given cosmology and halo mass definition, by enforcing the consistency conditions that: (i) at any time $t$ all the DM in the Universe is locked in halos of different masses, and (ii) the mass $M$ of a halo must be equal to the volume-integral of its density profile times the squared radius (see Sec.~\ref{density} for derivation of such a density profile). Specifically, if we write those two functions in the form
\beq
\delta(t,\ti)= r_\delta(t)\,\delta\cc\F(t)\frac{D(\ti)}{D(t)}
\label{deltat}
\eeq
\vspace{-10pt}
\beq
\sigma_0(R,t,\ti)=r_\sigma(M,t)\,\sigma_0\F(M,t)\frac{D(\ti)}{D(t)}\,,
\label{rmt}
\eeq
where $\delta\cc\F(t)$ is the linearly extrapolated density contrast for top-hat spherical collapse at $t$, $a(t)$ is the cosmic scale factor, $D(t)$ is the linear growth factor,\footnote{In the Einstein-de Sitter cosmology, $\delta\cc\F(t)=3(12\pi)^{2/3}/20=1.686$ and $D(t)=a(t)$; see e.g. \citep{H00} for other cosmologies.} $\sigma_0(R,t,\ti)$ is the Gaussian 0th-order spectral moment at $\ti$ on the scale $R$ corresponding to the mass $M$ at $t$ and $\sigma_0\F(M,t)$ is its top-hat counterpart at $t$, the functions $r_\delta$ and $r_\sigma$ appear to be well fitted in all cosmologies and halo mass definitions analysed by the analytic expressions
\beqa 
r_\delta(t)\approx \frac{a^{d{\cal D}(t)}(t)}{D(t)}
\,~~~~~~~~~~~~~~~~~~~~~~~~~~~~~~~~~~~~~~~~~~~~~~~~~~\nonumber\\
{\cal D}(t)=1-d_0a^{0.435/a(t)}(t)\phantom{\frac{}{}}~~~~~~~~~~~~~~~~~~~~~~~~~~~~~~~~~~~~~~\label{cc}\\
r_\sigma(M,t)\approx 1+r_\delta(t){\cal S}(t)\nu\F(M,t)~~~~~~~~~~~~~~~~~~~~~~~~~~~~~
\nonumber\\
{\cal S}(t)=s_0\!+\!s_1a(t)\!+\!\log\left[\frac{a^{s_2}(t)}{1\!+\!a(t)/A}\right],~~~~~~~~~~~~~~~~~~~~~~~~
\label{rs}
\eeqa
with $\nu\F(M,t)\equiv\delta\F(t,\ti)/\sigma\F_0(M,\ti)=\delta\F\cc(t)/\sigma\F_0(M,t)$ equal to the peak height in top-hat spherical collapse. See Table \ref{T1} for the values of coefficients $d$, $s_0$, $s_1$, $s_2$, and $A$ for several cosmologies (Table \ref{T2}) and halo mass definitions of interest. Note that while $R$ increases with increasing $M$, $\delta$ decreases with increasing $t$. We remark that the analytic fitting function $r_\delta(t)$ given by equation (\ref{cc}) only holds up to the present time; its extrapolation to larger times should be taken with caution (see Sec.~\ref{assembly}). When using virial masses, $M\vir$, $r_\sigma$ is a function of $M$ alone, $r_\sigma(M)= 1+c/\sigma_0\F(M)$, where $\sigma_0\F(M)$ is the top-hat 0th order spectral moment at the present time and $c=0.14$ and 0.10 in the {\it WMAP7} and {\it Planck14} cosmologies, respectively (Paper I).

\begin{table}
\begin{center}
\caption{Coefficients in the halo-peak relations.}
\begin{tabular}{ccccccccccc}
\hline \hline
Cosmol.\! & \!Mass$^*\!$ & $d$ & $10d_0\!\!$ & $10^{2}s_0\!\!$ & $10^{2}s_1\!\!$ & $10^{2}s_2\!\!$ & $A$ \\ 
\hline
\multirow{2}{*}
{WMAP7}\! & $M\vir\!$ & 1.06 & 3.0 & 4.22 & 3.75 & 3.18 & 25.7\\ 
   & $M_{200}\!$ & 1.06 & 3.0 & 1.48 & 6.30 & 1.32 & 12.4\\ 
\multirow{2}{*}
{Planck14}\! & $M\vir\!$ & 0.93 & 0.0 &  2.26 & 6.10 & 1.56 & 11.7 \\ 
   & $M_{200}\!$ &  0.93 & 0.0 & 3.41 & 6.84 & 2.39 & 6.87 \\
\hline
\end{tabular}
\end{center}
\label{T1}
$^*M\vir$ and $M_{200}$ are the masses inside the region with a mean inner density equal to $\Delta\vir(t)$ \citep{bn98} times the mean cosmic density, and 200 times the critical cosmic density, respectively.\\
\end{table}

\begin{table}
\begin{center}
\caption{Cosmological Parameters.}
\begin{tabular}{ccccccc}
\hline \hline 
Cosmology& $\Omega_\Lambda$ & $\Omega_{\rm m}$ & $h$ &
$n_{\rm s}$ & $\sigma_8$ & $\Omega_b$\\ 
\hline 
WMAP7$^{a}$ & 0.73 & 0.27 & 0.70 & 0.95 & 0.81 & 0.045\\
Planck14$^{b}$ & 0.68 & 0.32 & 0.67 & 0.96 & 0.83 & 0.049\\ 
\hline
\end{tabular}
\label{T2}
\end{center}
$^{a}$ \citet{Kea11}.\\
$^{b}$ \citet{P14}.
\end{table}

Strictly speaking, some peaks with $\delta$ at $R$ are nested into other peaks with the same $\delta$ at a larger scale, so they are actually captured by the more massive halo associated with the host peak before achieving full collapse and become subhalos instead of halos at $t$. Therefore, equations (\ref{deltat}) and (\ref{rmt}) do not define a {\it one-to-one} correspondence between halos and peaks. This means that the abundance of peaks with $\delta$ and $R$ at $\ti$ must be corrected for nesting in order to obtain the right mass function of halos at $t$ (Paper I). But in the present Paper we are not concerned with the peak number density, but with their average curvature, ellipticity, and prolateness, and these properties depend much more strongly on the background density of the peak (i.e. the density contrast at the same point at a larger scale) than on its possible location within another peak (i.e. at a different point with the same $\delta$ at a larger scale). Therefore, when calculating these properties, we will ignore the effects of their possible nesting in front of those of their background density.

\section{Peak Trajectory and Halo Density Profile}
\label{density}

Given the halo-peak correspondence (eq.~[\ref{deltat}]-[\ref{rmt}]), the equality
\begin{equation}
\frac{\partial\delta({\bf r},R)}{\partial R}=R\nabla^{2}\delta({\bf r},R)\equiv-x({\bf r},R)
\sigma_{2}(R)R
\label{Gau}
\end{equation}
fulfilled by the density field in {\it Gaussian smoothing} allows one to identify the peaks (possibly at slightly different points) ${\bf r}$ that trace the same evolving halo when the scale and the density contrast are varied accordingly in the $(\delta,R)$ plane at $\ti$ \citep{MSS95}. 

Specifically, when a halo accretes, the associated peak follows a continuous $\delta(R)$ trajectory, which is only interrupted when the halo undergoes a major merger.\footnote{Then, a new peak appears with the same $\delta$ but a substantially larger scale.} As shown next, the mean continuous peak trajectory traced by {\it purely} accreting halos with $M_0$ at $t_0$ ($t_0$ is not necessarily the present time) determines their average density profile. Certainly, halos also suffer major mergers, but, as mentioned in Section \ref{intro}, the density profile of halos with $M_0$ at $t_0$ does not depend on their merging history, so we can assume they evolve by pure accretion in order to derive their average density profile (and any other internal property; see Sec.~\ref{assembly}).

\subsection{Unconstrained halos}\label{density1}

According to equation (\ref{Gau}), the {\it mean} trajectory $\delta(R)$, solution of the differential equation 
\begin{equation}
\frac{\der \delta}{\der R}=-\lav x\rav [R,\delta(R)]\sigma_{2}(R)R,
\label{eq}
\end{equation}
with the boundary condition $\delta_0$ at $R_0$ corresponding to halos with $M_0$ at $t_0$, traces their average mass growth by accretion. In equation (\ref{eq}), $\lav x\rav(R,\delta)$ is the mean curvature of the peak at the intermediate point $\delta$ at $R$, given by \citep{MSS95}
\beq
\lav x\rav(R,\nu)=\frac{G_1(\gamma,\gamma\nu)}{G_0(\gamma,\gamma\nu)},
\eeq
where $G_i$ is the $i$th moment of $x$ for the $x$-PDF (BBKS)
\beqa
G_i(\gamma,\gamma\nu)\!=\!\int_0^\infty\! \der x\,x^i\,F(x) \frac{{\rm e}^{-\frac{(x-\gamma\nu)^2}{2(1\!-\!\gamma^2)}}}{[2\pi(1-\gamma^2)]^{1/2}}~~~~~~~~~~~~\label{hat}\\
F(x)\!\equiv \!\frac{(x^3-x)\left\{{\rm erf}\left[\!\left(\frac{5}{2}\right)^{\frac{1}{2}} x\!\right]\!+{\rm erf}\left[\!\left(\frac{5}{2}\right)^{\frac{1}{2}}\frac{x}{2}\!\right]\right\}}{2\!+\!\left(\frac{2}{5\pi}\right)^{\frac{1}{2}}\!\!\left[\!\left(\frac{31x^2}{4}+\frac{8}{5}\right){\rm e}^{-\frac{5x^2}{8}}\!+\!\left(\frac{x^2}{2}-\frac{8}{5}\right)
{\rm e}^{-\frac{5x^2}{2}}\!\right]}\!,~~
\eeqa
$x_\ast\equiv \gamma\nu$ and $\gamma\equiv \sigma^{2}_{1}/(\sigma_{0}\sigma_{2})$, being $\sigma_{\rm j}$ the $j$-th spectral moment. In  the case of power-law power spectra of index $n$, $\gamma$ is constant and equal to $[(n+3)/(n+5)]^{1/2}$, while in the case of the Cold Dark Matter (CDM) spectrum, locally close to a power-law with index $n\approx -1.75$ in the range of galactic mass halos, we have $\gamma\approx 0.62$.

Equation (\ref{eq}) shows that the mean curvature of peaks at $R$ determines the accretion rate of the corresponding halos, and  that the mean peak trajectory $\delta(R)$ traces the average accretion history of halos with $M_0$ at $t_0$. Moreover, since accreting halos grow inside-out \citep{Sea12a,SM19}, their accretion history automatically sets their density profile, so the mean peak trajectory $\delta(R)$ determines the average density profile of those halos with $M_0$ at $t_0$. 

Specifically, as shown in \citet{SM19}, the peak trajectory is the convolution with a Gaussian window of the peak density profile. Thus, one can infer the average density profile of halos with $M_0$ at $t_0$ by deconvolving the mean peak trajectory solution of equation (\ref{eq}) and monitoring their monolithic ellipsoidal collapse and virialization (taking into account that both processes preserve the radial mapping of the initial mass distribution; \citealt{Sea12a}).\footnote{The same procedure could be applied to individual halos, though the random peak trajectory of a specific halo is unknown, in general.} The resulting density profile is of the NFW \citep{NFW95} or the Einasto \citet{E65} form with a concentration that scales with halo mass as found in simulations over more than 20 orders of magnitude \citep{Sea23}. Moreover, similar procedures using the ellipticity and prolateness of peaks instead of their curvature lead to the average shape and kinematics of halos (see \citealt{Sea12b,SM19}). 

Interestingly, all these derivations can be applied not only to unconstrained halos (and peaks), but also to halos (peaks) constrained to lie in different backgrounds. Since the curvature, ellipticity, and prolateness of constrained peaks are different from those of unconstrained ones, the properties of halos lying in different backgrounds will also differ from those of unconstrained halos, which could explain the secondary bias. To check this possibility and find the strength of the effect according to the background height, we should compare the different properties obtained for both kinds of objects with varying backgrounds and halo masses. But that would be a very laborious task and would not clarify the physical reason for the results we would obtain. Fortunately, there is an alternative, fully analytic way to do this that only makes use of the change in the mean curvature between unconstrained and constrained peaks. 


\subsection{Halos Constrained to Lie in a Background}
\label{background}

Let us now turn to halos with $M_0$ at $t_0$ constrained to lie in a background with matter density contrast $\delta\m(t_0)$. To distinguish the properties of these constrained halos from those of unconstrained ones, we will hereafter denote the former with index ``co''. This includes the properties of the corresponding seeds: peaks with density contrast $\delta\co_0=\delta_0$ at $R_0$ lying in a background with (matter) density contrast $\delta\m(\ti)=\delta\m(t_0) D(\ti)/D(t_0)$ at a scale $R\m$ substantially larger than $R_0$. 

Even though peaks along a trajectory $\delta\co(R)$ tracing the evolution of accreting halos slightly slosh around a given location when $R$ increases, since the scale of the background at $\ti$ is substantially larger than $R$, they necessarily keep lying on the same background. As a consequence, the mean trajectory of peaks corresponding to halos constrained to lie on the background is now the solution of the differential equation 
\begin{equation}
\frac{\der \delta\co}{\der R}=-\lav x\rav [R,\delta\co(R)|R\m,\delta\m]\sigma_2(R)R,
\label{cond}
\end{equation}
where $\lav x\rav [R,\delta\co(R)|R\m,\delta\m]$ is the mean curvature of peaks with $\delta\co$ at $R$ lying in a background with $\delta\m$ at scale $R\m$. 

As shown in \citet{MSS96}, this conditional mean curvature takes exactly the same form as the unconditional one, equation (\ref{hat}), but with $\gamma\nu$ replaced by $\tilde\gamma\tilde\nu$, given by (BBKS)
\begin{equation}
\tilde\gamma=\gamma\left[1+\epsilon^{2}\frac{(1-r_{1})^{2}}{(1-\epsilon^{2})}\right]^{1/2}
\end{equation}
\begin{equation}
\tilde\nu=\frac{\gamma}{\tilde\gamma}\frac{(1-r_{1})}{(1-\epsilon^{2})}\left[\nu\frac{(1-\epsilon^{2}r_{1})}{(1-r_{1})}-\epsilon\nu\m\right],
\end{equation}
where $\nu\m\equiv \delta_m/\sigma_{\rm 0m}$, being $\sigma_{j{\rm m}}\equiv \sigma_j(R\m)$,
\begin{equation}
\epsilon\equiv\frac{\sigma^{2}_{\rm 0h}}{\sigma_{0}\sigma_{\rm 0m}}, \hspace{1cm} r_{1}\equiv\frac{\sigma^{2}_{\rm 1h}\sigma^{2}_{0}}{\sigma^{2}_{0h}\sigma^{2}_{1}}, \nonumber
\end{equation}
with $\sigma_{j{\rm h}}(R,R\m)$ defined as $\sigma_{j}$ but with $R$ replaced by the squared mean scale $R\h\equiv [(R^{2}+R\m^{2})/2]^{1/2}$. Note that, in the limit $R\m\rightarrow\infty$, $\epsilon$ vanishes, and we have $\tilde\gamma=\gamma$ and $\tilde\nu=\nu$, so the mean curvature of constrained peaks becomes equal to that of unconstrained ones, as expected. 

With those expressions, the quantity $\tilde \gamma \tilde \nu$ takes the explicit form
\begin{equation}
\tilde \gamma\tilde \nu =\frac{\delta(\sigma^{2}_{\rm 0m}\sigma^{2}_{1}-\sigma^{2}_{\rm 0h}\sigma^{2}_{\rm 1h})-\delta f\m (\sigma^{2}_{\rm 0h}\sigma^{2}_{1}-\sigma^{2}_{\rm 1h}\sigma^{2}_{0})}{\sigma_{2}(\sigma^{2}_{0}\sigma^{2}_{\rm 0m}-\sigma^{4}_{\rm 0h})}.
\label{eq:20}
\end{equation}
At this point, it is convenient to adopt the same approximation used in Paper I. Taking into account that $R\m$ is substantially larger than  $R$ (say, $R\m\ga 3R$), $R^2$ can be neglected in front of $R^2\m$, so expression (\ref{eq:20}) becomes $\tilde \gamma\tilde \nu=\gamma\nu_{\rm e}$, where $\nu_{\rm e}=\delta_{\rm e}/\sigma_0$ is defined in terms of the effective density contrast $\delta_{\rm e}=\delta-q(R\m)\delta\m$, being $q(R\m)\equiv \sigma_0^2(R\m/\sqrt{2})/\sigma_0^2(R\m)$. In the case of power-law power spectra, $q$ is constant and equal to $2^{(n+3)/2}$ while, in the case of the CDM spectrum locally close to a power-law, the same expression approximately holds with the effective value of $n$ corresponding to the scale $R\m$. From now on, we adopt the effective constant value $q\approx 1.6$ shown in Paper I to yield very good results for galactic halo backgrounds.

Thus, the mean average curvature of peaks with $\delta\co$ at $R$ lying in a background $\delta\m$, $\lav x\rav [R,\delta\co(R)|R\m,\delta\m]$, is very nearly equal to the average curvature of unconstrained peaks with effective density contrast $\delta\co\e=\delta\co-q\delta\m$ at $R$. Consequently, the mean trajectory of constrained peaks is very nearly given by the solution of the equation  
\begin{equation}
\frac{\der \delta\co}{\der R}=-\lav x\rav [R,\delta\co(R)-q\delta\m]\sigma_2(R)R,
\label{condcon}
\end{equation}
for the boundary condition $\delta\co_0=\delta_0$ (both kinds of peaks collapse at the same time $t_0$) at $R_0$.
Since $q\delta\m$ is constant, we can re-write equation (\ref{condcon}) as
\begin{equation}
\frac{\der \delta\e}{\der R}=-\lav x\rav [R,\delta\e(R)]\sigma_2(R)R,
\label{condconbis}
\end{equation}
showing that the mean trajectory $\delta\co(R)$ of constrained peaks coincides with the mean trajectory $\delta\e(R)$ of some equivalent unconstrained ones, with boundary condition $\delta_{\rm e0}=\delta_0-q\delta\m$ at $R_0$. 

Given the relation between the mean peak trajectory and the average halo density profile, we are led to the conclusion that {\it the average density profile of halos with $M_0$ at $t_0$ lying in a background $\delta\m(t_0)$ is equal to the average density profile of the associated unconstrained halos} (from now on simply the unconstrained halos) {\it with $M_0$ at $t_{\rm e0}\equiv t(\delta_0-q\delta\m)$ $(t_{\rm e0}>t_0)$}, being $t(\delta)$ the inverse of the function $\delta(t)$ given by equation (\ref{deltat}). Note that, according to equation (\ref{eq}), this also means that halos lying in different backgrounds have distinct accretion rates (not to  mix up with distinct merger rates), in agreement with the results of simulations \citep{Lea17,Cea20}.

\section{Secondary Bias}
\label{assembly}

In Section \ref{density} we saw that halos with $M_0$ at $t_0$ lying in different backgrounds arise from peaks with $\delta_0$ at $R_0$ having different average curvatures, and hence, tracing different $\delta(R)$ trajectories. In the present Section we will see that, as a consequence, the corresponding haloes have different formation times ($z_{\rm f}$), concentrations ($c$), peak velocities ($V\maxi$), subhalo abundances ($N_{\rm s}$), kinematic profiles (velocity dispersion $\sigma$ and anisotropy $\beta$), triaxial shapes (ellipticity $e$ and prolateness $p$), and spins ($\lambda$) as found in simulations (\citealt{ST04,Gea05,Hea06,Wech06,GW07,Lea17,Mea18,SP19,Cea20}). 

\begin{figure*}
\centering
 {\includegraphics[scale=1.20,bb= 80 0 210 200]{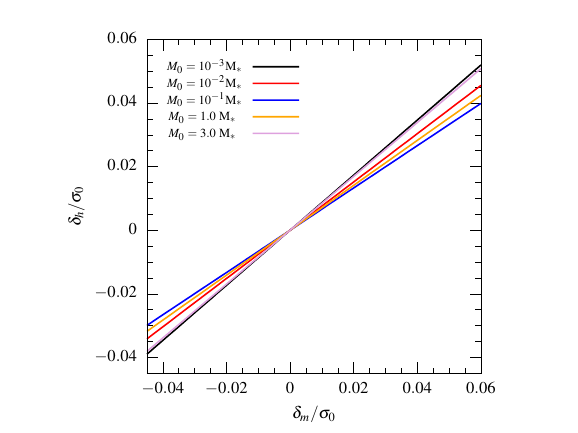}
 \includegraphics[scale=1.20, bb= 5 0 210 200]{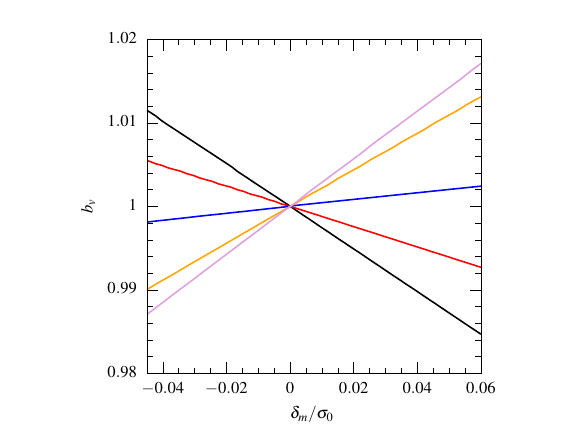}
 \caption{Two different measures of halo clustering: halo overdensity (left panel) and the bias parameter $b_{v}$ defined in the text (right panel), as a function of the background matter overdensity $\delta\m(t_0)$ (for simplicity, we drop the argument $t_0$) for several virial masses $M_0$ in the {\it Plancl14} cosmology at $z_0=0.5$.}}
\label{bp}
\end{figure*}

One indicator, used in the pioneer work by \citet{ST04},  that a property $P$ is biased is that its median value $v\co/v$ in halos with $M_0$ at $t_0$ lying in regions with different halo overdensities $\delta\h\approx b\, \delta\m(t_0)$, where $b$ is the linear halo bias (Paper I), scaled to the median value $v$ in unconstrained halos depends on $\delta\m(t_0)$. To show this in the case of properties directly related to the halo density profile we will use that their median values coincide with the values of those properties in a halo with the average density profile. Indeed, the proof given in \citep{Sea23} for the concentration automatically translates to any property of that kind, provided only it is a monotonic function of the concentration. For the remaining properties, namely the halo shape and spin, we will directly use the corresponding median values.

Another indicator of a biased property, introduced by \citet{Wech06} and used in most studies of secondary bias, is that the quantity $b_{v}\equiv \sqrt{\xi\h(r;v\co)/\xi\h(r)}$, where $\xi\h(r;v\co)$ and $\xi\h(r)$ are the autocorrelations at the scale $M_0$ of halos with $M_0$ and a specific value $v\co$ of the property and with $M_0$ alone, respectively, depends on $v\co$ (and differs from unity). As readily seen by replacing $\xi\h(r)$ by $b^2\xi(r)$ in the definition of $b_{v}$, this parameter is nothing but $b\co/b$, where $b\co$ is the linear bias of halos with $M_0$ and $v\co$, given by the same expression as the plain linear halo bias $b$ derived in Paper I,\footnote{As shown in Paper I, the linear bias $b$ calculated by means of CUSP using Gaussian smoothing coincides with those found in simulations using top-hat smoothing.} but with $\lav x\rav(R_0,\delta_0)$ of peaks with $\delta_0$ at $R_0$ replaced by $\lav x\rav(R_0,\delta_{\rm e0})$ with $\delta_{\rm e0}=\delta_0-q\delta\m$. 

Since each of these indicators has its own interest (the simplicity of the former case, and the frequent use of the latter), we will analyze both. It is important to realize that the median $v\co/v$ vs. $\delta\h$ and the $b_{v}$ vs. $v\co/v$ relations result from the parametrization through $\delta\m(t_0)$ of two more fundamental relations: (i) the $v\co/v$ vs. $\delta\m(t_0)$ relations, and (ii) the $\delta_{v}$ or $\delta\co$ vs. $\delta\m(t_0)$ relations. Relations (i) for the different properties $P$ will be derived below, while the two $P$-independent (see their definition above) relations (ii) are shown in Figure 1.

In this Figure we see that, while $\delta\h$ is linear with $\delta\m(t_0)$ with relatively similar positive slopes $b$ for all relevant $M_0$, $b_{v}$ is also linear with $\delta\m(t_0)$ but with a slope that markedly depends on $M_0$, with the opposite sign for $M_0$ lower or higher than a few $10^{-1}M_\star$ (slightly less than $M_\star$) for virial masses in the {\it Plank14} ({\it WMAP7}) cosmology, where $M_\star$ is the cosmology-dependent typical mass of top-hat spherical collapse (solution of the equation $\sigma_0\F(M_\star,t_0)=\delta\cc\F(t_0)$). The behavior of both relations is well-understood: the larger the background $\delta\m(t_0)$, the higher the overdensity $\delta\h$ of halos of mass $M_0$ lying in it, while the behavior of $b_{v}$ with $\delta\m$ arises from the value of $\lav x\rav(R,\delta_0-q\delta\m)$, being $\lav x\rav(R,\delta)$ a monotonically increasing or decreasing function of $\delta\m$ depending on the value of $M_0$. This means that the behavior of $b_{v}$ has nothing to do with the more or less marked clustering of halos with fixed values of any property $P$ when $M_0$ is varied. It is simply due to the behavior of this clustering indicator with $\delta\m(t_0)$.

\begin{figure}
\includegraphics[scale=1.30,bb= 45 0 210 200]{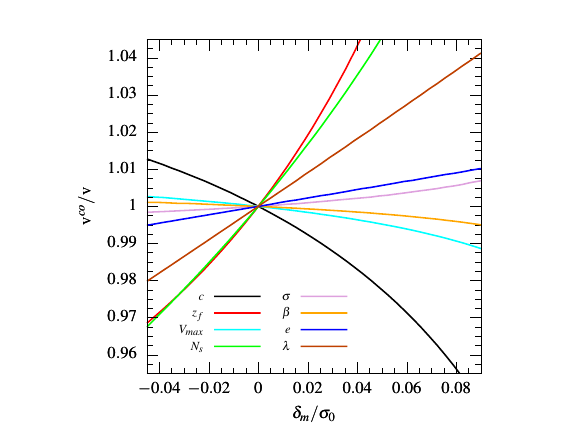}
 \caption{Comparison between the predicted bias in the different properties analyzed here for halos with virial mass $M_0=M_\ast$ in the {\it Planck14} cosmology.
 (A color version is available in the online journal.)}
\label{all}
\end{figure}

Relations (ii) combined with relations (i) derived below, approximately satisfying
\beq
\frac{v\co}{v}\approx 1+\frac{\der \frac{v\co}{v}}{\der \delta\m}\Big|_{\delta\m=0} \delta\m+ \frac{1}{2}\frac{\der^2 \frac{v\co}{v}}{\der \delta\m^2}\Big|_{\delta\m=0} \delta\m^2 
\eeq
in small enough $\delta\m$ ranges (see Figure \ref{all}), cause the $v\co/v$ vs. $\delta\h$ and $b_{v}$ vs. $v\co/v$ plots, with the $v\co/v$ limits covering the same $\delta\m$ range, to be very similar if not identical for all properties $P$. Thus, we will only show them for two halo properties: the concentration and the formation time. This is enough to illustrate the great similarity in the plots of all properties except for their decreasing or increasing trends with $\delta\m$, and to facilitate the comparison of our predictions to the results of simulations in these best studied cases. 

All Figures shown in this Paper are for $M\vir$ masses, the {\it Plank14} cosmology, and redshift $z_0=0.5$. The reason for not adopting $z_0=0$ is that the collapse time of the equivalent unconstrained peaks, $t\e=t(\delta_0-q\delta\m)$, exceeds in this case the present time and goes beyond the interval used to fit the function $r_\delta(t)$ (eq.~[\ref{deltat}]). Nevertheless, as shown in Paper I, the linear biases $b$ and $b\co$ are nearly universal when using $M\vir$ masses (and expressing them as functions of $\nu\F$), so their ratio $b_{v}$ is nearly universal too (regardless of how they are expressed). In addition, overdensities are scaled to $\sigma_0$ and halo masses to $M_\star$ so that all the plots are essentially independent of cosmology (for the suited value of $M_\ast$) and redshift. In fact, numerical studies also use  snapshots at different cosmic times (up to a redshift as large as $z_0=3$) so as to enhance the resolution of simulations (e.g. \citealt{Wech06,GW07}) and scale similarly all quantities with the same purpose.

\begin{table*}
\begin{center}
\caption{Coefficients of the mass-concentration relation for NFW halos.}
\vspace{15pt}
\begin{tabular}{cccccccccc}
\hline \hline Cosmology & Mass & $r_{\rm c\ast}$ (Mpc) & $M_{\rm c\ast}$ (\modotc) & $\tau_{\rm c\ast}$ & $t_1$ & $t_2$ & $t_3$ \\ \hline 
\multirow{2}{*}{WMAP7} & $M\vir$ & $9.46\times 10^{-5}$ & $1.00\times 10^5$  & 0.325 & 0.183 & .0145 & $-0.187$ \\ 
      & $M_{200}$ & $9.75\times 10^{-5}$ & $1.00\times 10^5$  & 0.317 & 0.199 & .0134 & $-0.121$ \\ 
      \multirow{2}{*}{Planck14} & $M\vir$  & $8.04\times 10^{-5}$ & $1.00\times 10^5$  & 0.280 & 0.382 & .00854 & $-0.110$ \\
      & $M_{200}$ & $8.59\times 10^{-5}$ & $1.00\times 10^5$  & 0.314 & 0.219 & .0134 & $-0.131$ \\ 
\hline
\end{tabular}\hspace{60pt}
\label{T3}
\end{center}
\vspace{-5pt}
\end{table*}

One final remark is in order. To facilitate obtaining fully analytic expressions for the $v\co/v$ vs. $\delta\m(t_0)$ relations we will take advantage that the average density profile of unconstrained halos with $M_0$ at $t_0$ is well approximated \citep{Sea23} by the NFW analytic profile \citep{NFW95}, 
\beq
  \rho(r)=\rhos\frac{4\rs^3}{r\left(r+\rs\right)^2},
\label{NFW}
\eeq
where $\rs$ and $\rhos$ are the so-called core radius and characteristic density, respectively. Another useful quantity related to this profile is the mass inside the core radius $r\cc$, $M\cc=16\pi r^3\cc  f(1) \rho\cc$, where $f(x)=ln(1+x)-{x}/(1+x)$, related to the total mass $M$ through 
\beq
M_{\rm c}=\frac{f(1)}{f(c)}M_0,
\label{car}
\eeq
being $c$ the halo concentration. The price we must pay for this is that, since the fit to the analytic NFW profile is not perfect, the best fitting values of $\rs$ slightly vary with time even though the real core radius is kept unchanged when the density profile grows inside-out \citep{SMS05}. Since this spurious effect is more marked for low mass halos, our predictions are only shown for halos with masses $M_0> 10^{-3}M_\ast$.

\subsection{Concentration}\label{concentration}

The concentration $c$ of a halo is defined as the total radius over the core radius, $c=r_0/r\cc$. Since the virial radius $r_0$ of all halos with $M_0$ at $t_0$ is the same, 
\beq
r_0= \left[\frac{3M_0}{4\pi\Delta\vir(t_0)\bar\rho(t_0)}\right]^{1/3},
\label{vir}
\eeq
the concentration $c\co$ of constrained halos with $M_0$ at $t_0$ can only differ from the concentration $c$ of unconstrained ones through their distinct core radius.

\begin{figure*}
\centering
 {\includegraphics[scale=1.20,bb= 80 0 210 200]{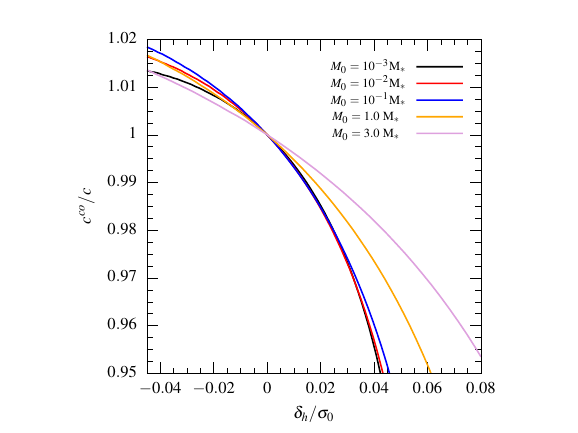}
 \includegraphics[scale=1.20, bb= 5 0 210 200]{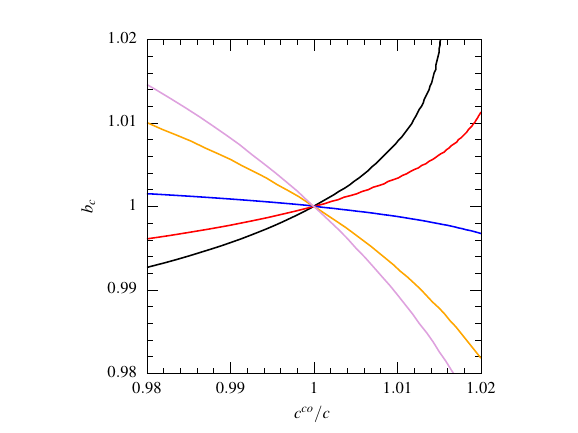}
 \caption{Left panel: Ratio of median concentrations of constrained and unconstrained halos, $c\co/c$, as a function of the halo overdensity $\delta\h$ scaled to the rms matter density contrast $\sigma_0$ for halos of several virial masses in the {\it Planck14} cosmology. Right panel: Bias parameter $b\cc$ (i.e. $b_{v}$ for $v=c$) commonly used in numerical studies of the secondary bias as a function of $c\co/c$.
 (A color version is available in the online journal.)}}
\label{bconc}
\end{figure*}

As shown in Section \ref{background}, the density contrast of the equivalent unconstrained peaks leading to the average density profile of constrained halos with $M_0$ at $t_0$ (or redshift $z_0$) is $\delta_{\rm e0}=\delta_0-q\delta\m$, which is smaller than $\delta_0$. Consequently, the collapse time of the equivalent unconstrained halos, $t\e=t(\delta_0-q\delta\m)$, is larger than $t_0$ (and the corresponding redshift $z\e\equiv z(t\e)$ smaller than $z_0$). As mentioned, CUSP allows one to compute the density profile of unconstrained halos and, from it, their median concentration. Specifically, unconstrained halos with mass $M_0$ at $t_0$ (or $z_0$), with the characteristic mass $M_{\rm c}$ (eq.~[\ref{car}]) and total radius $r_0$ (eq.~[\ref{vir}]), have a median concentration $c$ satisfying \citep{Sea23}  
\beq 
c=\frac{r_0}{r_{\rm c\ast}}\left(\frac{M_{\rm c}}{M_{\rm c\ast}}\right)^{-\tau(M_{\rm c},t_0)},
\label{Mcz}
\eeq
where
\beq
\tau(M_{\rm c},t_0)=\tau_{\rm c\ast}\left[1 + t_1\left(\frac{M_{\rm c}}{M_{\rm c\ast}}\right)^{t_2}(1+z_0)^{t_3}\right],
\label{tau}\\
\eeq
with coefficients $\tau_{\rm c\ast}$, $r_{\rm c\ast}$, $M_{\rm c\ast}$, $t_1$, $t_2$, and $t_3$ given in Table \ref{T3}.\footnote{Expressions (\ref{Mcz})-(\ref{tau}) coincide with those given in \citet{Sea23}, but they are presented in a more compact way, with the coefficients $t_i$ redefined accordingly.} 

This relation also holds, of course, for the equivalent {\it unconstrained} halos collapsing at $t\e$ (redshift $z_{\rm e}$). Given that the equivalent unconstrained halos grow inside-out, their average density profile at $t_0$ is also of the NFW form with the same core radius and a smaller concentration $c\co$ because the total radius is smaller, $r\co_0< r\e$. Thus, using the redshift dependence of the median concentration of unconstrained halos, $c\propto (1+z)^{-1}$ \citep{Sea23}, we have 
\beq
c\co=c\e\,\frac{1+z\e}{1+z_0}.
\label{proxi}
\eeq

As mentioned, the higher the background density $\delta_{\rm m}(t_0)$, the lower $z\e$, so equation (\ref{proxi}) tells that the smaller $c\co$ compared to $c\e$. But $c\co$ compared to $c$ also depends on $M_0$ because $c\e$ does. Specifically, taking into account that the unconstrained halos with concentrations $c\e$ and $c$ satisfy equations (\ref{Mcz})-(\ref{tau}) for the total radii $r\e$ and $r_0$, characteristic masses $M_{\rm ce}$ and $M_{\rm c}$, and times $t\e$ and $t_0$, respectively, equation (\ref{proxi}) leads to 
\beq
\frac{c\co }{c}\!=\frac{(M_{\rm c}/M_{\rm c\ast})^{\tau(M_{\rm c},t_0)}}{(M_{\rm ce}/M_{\rm c\ast})^{\tau(M_{\rm ce},t\e)}} \left[\frac{\Delta\vir(t_0)}{\Delta\vir(t_{\rm e})}\right]^{1/3}.
\label{proxi3}
\eeq

\begin{figure*}
\centering
{\includegraphics[scale=1.20,bb= 80 0 210 200]{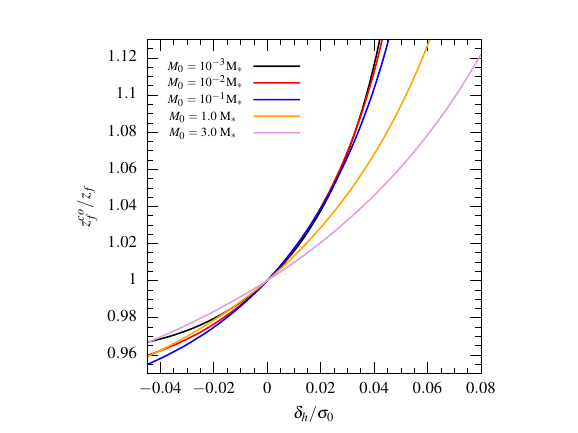}
\includegraphics[scale=1.20, bb= 5 0 210 200]{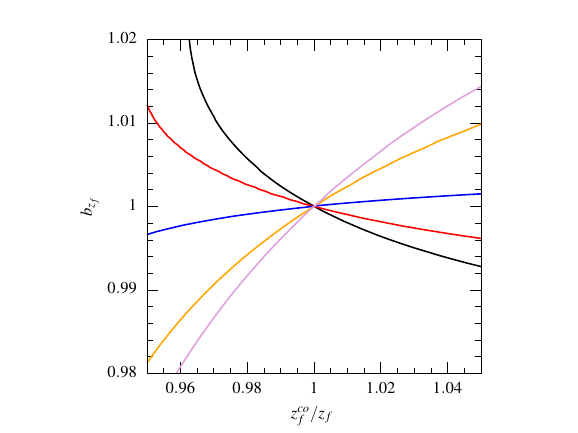}
 \caption{Same as Figure 3 but for the halo formation time. The limits in the abscissa of the left panel correspond to those of in the left panel of that Figure.
 (A color version is available in the online journal.)}}
 \label{bzf}
\end{figure*}

The predicted dependence of the typical concentration $c\co$ of halos with $M_0$ at $t_0$ lying in an overdensity region $\delta\h$ of halos with $M_0$ is shown in the left panel of Figure 3, while the dependence of $b\cc$ (i.e. $b_{v}$ for $v=c$) for halos with $M_0$ and $c\co$ at $t_0$ is shown in the right panel. In the latter we see that, as found in simulations, for low-mass halos the higher the concentration, the higher $b\cc\equiv b\co/b$, while the opposite is true for high-mass halos (compare this plot with Fig.~4 of \citealt{Wech06}). However, as seen in the left panel, the typical value of $c\co/c$ is monotonically decreasing with increasing $\delta\h$ regardless of the halo mass. This clearly shows that, for the reasons explained at the beginning of this Section, the reversal of the $b\cc$ vs. $c\co/c$ relation from low to high masses has no physical relevance.

\subsection{Formation Time}\label{formation}

The formation time $t_{\rm f}$ (or formation redshift $z_{\rm f}$) of a halo with $M_0$ at $t_0$ is defined as the time (redshift) at which the halo reached half its current mass. Since the higher the concentration of a halo, the larger its mass fraction at small radii, the earlier they have also formed.  

More specifically, the half-mass radius $r_{\rm h}$ is related to the formation time, $t_{\rm f}$ of unconstrained halos with $M_0$ at $t_0$ through (see eq.~[\ref{vir}])
\beq
r_{\rm h}= \left[\frac{3M_0/2}{4\pi\Delta\vir(t_{\rm f})\bar\rho(t_{\rm f})}\right]^{1/3}.
\label{half}
\eeq
On the other hand, in NFW halos with mass $M_0$, radius $r_0$, and concentration $c$, the mass inside the radius $r$ satisfies the relation
\beq
M(r)=M_0\,\frac{f(c\,r/r_0)}{f(c)}\,,
\label{new}
\eeq
implying that the radius $r_{\rm h}$ encompassing half their total mass is the solution of the implicit equation
\beq
\frac{f(c\,r_{\rm h}/r_0)}{f(c)}=0.5.
\label{half0}
\eeq
Therefore, equation (\ref{half0}) with $r_{\rm h}$ given by equation (\ref{half}) is an implicit equation for the formation time $t_{\rm f}$ of halos with $M_0$ at $t_0$. 

The previous relations also hold, of course, for the equivalent unconstrained halos (of the NFW form) with $M_0$ at $t_0$ by simply replacing $c$ by $c\co$. Consequently, taking into account that $\Delta\vir$ and $f$ are very smooth functions of their respective arguments, both implicit equations lead, after some algebra, to the following relation between the typical formation redshift $z\co_{\rm f}$ of constrained and unconstrained halos with $M_0$ at $t_0$ 
\beq
\frac{z\co_{\rm f}}{z_{\rm f}}\approx 1+\left(\,\frac{c}{c\co}-1\right)\frac{1+z_{\rm f}}{z_{\rm f}}.
\label{zf}
\eeq
Since the more strongly clustered halos, the lower their concentration, equation (\ref{zf}) tells that the larger their formation redshift. 

As can be seen in Figure \ref{all}, the dependence of the formation redshift with $\delta\m(t_0)$ is substantially steeper than that of the concentration, and shows the opposite trend. Yet, the plots in each panel of Figure 4 are identical (very similar in other cases below) to those of Figure 3 except for the opposite trend of the curves. In particular, in the right panel of Figure 4 we see the same reversal of the trend in the $b_{z_{\rm f}}$ vs. $z_{\rm f}\co/z_{\rm f}$ relation from low to high masses. We remark that constrained halos with very low masses (say, with $M_0<10^{-1}M_\star$) will form at such high redshifts that their mass at formation will often fall below the halo mass limit of simulations, so they will not be included in numerical studies of this bias. As a consequence, the reversal of the trend should be more difficult to detect for the formation redshift than for the concentration, which explains the doubts about that particular feature regarding the formation time found in the literature (e.g. \citealt{Jea07,GW07}).

\subsection{Peak Velocity}

The peak velocity $v_{\rm p}$ of a halo is the maximum value reached by its circular velocity profile $V(r)$ defined as 
\beq
V(r)=\left[\frac{GM(r)}{r}\right]^{1/2},
\label{v}
\eeq
where $M(r)$ is the mass inside $r$ and $G$ is the gravitational constant. Obviously, the more concentrated a halo of a given total mass, the steeper the profile $M(r)$, so the higher also the peak velocity. Let us put this in a more quantitative way. 

By differentiating equation (\ref{v}), we have that the radius $r\maxi$ marking the maximum circular velocity or peak velocity $V\maxi$ satisfies
\beq
\frac{V(r\maxi)}{2 r\maxi}\left[\frac{\der \ln M}{\der \ln r}\bigg|_{r\maxi}-1\right]=0,
\eeq
implying $4 \pi r\maxi^3 \rho(r_{\rm max})=M(r\maxi)$, which, in NFW halos with $M_0$ and $r_0$ at $t_0$, leads to
$r_0/(c\,r\maxi)=f^{-1/2}(c\,r\maxi/r_0)-1$. Plugging such an expression of $r\maxi$ into equation (\ref{v}) and using the NFW expression for $M(r\maxi)/r\maxi$, we obtain
\beq
V\maxi=\left[\frac{GM_0}{r_0} \frac{h(c\,r\maxi/r_0)}{h(c)}\right]^{1/2},
\eeq
where $h(x)=f(x)/x$. 

Replacing $c$ by $c\co$, we obtain the homologous expression for halos of the same mass $M_0$ at the same time $t_0$ constrained to lie in a background. And, from both relations, we arrive at the following ratio between the peak velocity $V\co\maxi$ in constrained halos and its counterpart $V\maxi$ in unconstrained ones,
\beq
\frac{V\co\maxi}{V\maxi}=\left[\frac{h(c)}{h(c\co)}\right]^{1/2},
\label{vp}
\eeq
where we have used that $h[(c\, r\maxi)\co/r_0]$ equals $h[(c\,r\maxi)/r_0]$ as $c\,r\maxi/r_0$ and $(c\,r\maxi)\co/r_0$ are solutions of the same implicit equation mentioned above.

%

In Figure \ref{all} we see that the $V\co_{\rm max}/V\maxi$ decreases with increasing $\delta\m(t_0)$ like $c\co/c$ though substantially less steeply. In this sense, even though the peak velocity is often used to evidence the concentration bias  because $V\maxi$ is simpler to measure than $c$ (e.g. \citealt{GW07,A08}), one must bear in mind that the former has a much less marked bias than the latter. What is instead a very good indicator of the concentration bias (though with the opposite trend) is the bias in the peak velocity radius since one is then led to $r\maxi\co/r\maxi=c/c\co$. 


\subsection{Subhalo Abundance}

The results of Section \ref{density} were obtained assuming purely accreting halos. This was justified because the properties of halos with $M_0$ at $t_0$ do not depend on their merger history. But, as mentioned, there is one exception: the properties related to substructure.\footnote{The reason for this is that subhalos suffer tidal stripping and dynamical friction, so their related properties do not directly arise from the collapse and virialization processes \citep{I,II,III}.} In particular, the subhalo abundance down to a fixed scaled subhalo mass, $M_{\rm s}/M_0$, $N_{\rm s}$, depends on the time $t_{\rm m}$ of the last major merger of the halo \citep{III}. Strictly speaking, $t_{\rm m}$ is not the same as the halo formation time $t_{\rm f}$. But, since halos essentially double their mass in major mergers (e.g. \citealt{Rea01}), we can take the latter as a good proxy for the former. 

The dependence on $t_{\rm f}$ of $N_{\rm s}$ is quite convoluted: apart from depending on the accretion rate of the host halo \citep{I}, it depends on its concentration determining the strength at which accreted subhalos are tidally stripped by the halo potential well \citep{II} and the halo merger history \citep{III}. However, $\log N_{\rm s}$ in the 20\% of unconstrained halos with $M_0$ at $t_0$ having suffered the last merger before $0.28t_0$ is $\approx 3/2$ times that of the 20\% of objects having suffered it after $0.76t_0$ \citep{III}. Thus, adopting the approximation that the merger of the two kinds of halos took place just at the time delimiting their respective intervals, we are led to 
\beq
\log (N_{\rm s}/N_{\rm s0})\sim -A \log (t_{\rm f}/t_0),
\label{relu}
\eeq
where $N_{\rm s}$ and $N_{\rm s0}$ are the subhalo abundances of unconstrained halos with $M_0$ at $t_0$ formed at $t_{\rm f}$ and at whatever time, respectively, and factor $A$ satisfies the equation $\log(3/2)=-A(\log 0.28-\log 0.76)$ is $A\sim 0.41$. 

Relation (\ref{relu}) also holds, of course, for the equivalent unconstrained halos with $M_0$ at $t_{\rm e0}=t(\delta_0-\delta\m)$, so, dividing both relations, we are led to  
\beq
\frac{N\co_{\rm s}}{N_{\rm s}}\approx  \left(\frac{t_{\rm f}\co t_0}{t_{\rm f}t_{\rm e0}}\right)^{-0.41}\frac{N_{\rm se}}{N_{\rm s0}},
\label{N}
\eeq
where $N_{\rm s}$ and $N_{\rm se}$ are the subhalo abundances (down to the same scaled subhalo mass) of unconstrained halos with  $M_0$ at the times $t_0$ and $t\e$, respectively. As shown in \citet{III}, the average subhalo abundance scales with the mass $M_0$ of halos (with their own typical concentration) as $N_{\rm s}\propto (M_0/10^{12}$\modotc$)^{0.08}$, with the same proportionality factor at any time $t_0$. Since both kinds of unconstrained halos have identical mass $M_0$, the last factor on the right of equation (\ref{N}) cancels and we arrive at
\beq
\frac{N\co_{\rm s}}{N_{\rm s}}\approx  \left(\frac{t_{\rm f}t_{\rm e0}}{t_{\rm f}\co t_0}\right)^{0.41}.
\label{Nbis}
\eeq
  
Given that the higher $\delta\m(t_0)$, the earlier halos form and the later they collapse, equation (\ref{Nbis}) tells that the larger their amount of subhalos, in agreement with the results of simulations. Both characteristic times are functions of concentration (Secs.~\ref{concentration} and \ref{formation}), so the subhalo abundance in halos with the average density profile coincides with its median value, like in all preceding properties. Figure 4 shows that the bias in the subhalo abundance is very similar to that in the formation redshift.


\subsection{Kinematics}

The velocity dispersion $\sigma(r)$ and anisotropy $\beta(r)$ profiles of haloes with $M_0$ at $t_0$ are related to the curvature, ellipticity, and prolateness of the corresponding peaks in a convoluted non-analytic way that involves, in addition, their triaxial shape \citep{Sea12b}. However, taking into account energy conservation and the gravitational origin of the velocity anisotropy, it was possible to prove the existence of two well-known universal relations found in simulations: one for the velocity dispersion profile  \citep{B85,TN01}
\beq
\sigma(r)\propto \rho^{1/3}(r)r^{1.875/3},
\label{sigma}
\eeq
with universal proportionality factor, and the other for the velocity anisotropy profile $\beta(r)$ \citep{HS06}
\beq
\beta(r) \approx -0.2\left(\frac{\der\ln\rho}{\der \ln r}+0.8\right).
\label{beta0}
\eeq
In other words, both kinematic profiles turn out to be fully determined by the density profile itself. The ultimate reason for this is well understood: as mentioned in Section \ref{correspondence}, the typical ellipticity and prolateness profiles of peaks are functions of the peak trajectory, which determines the density profile of the final halo \citep{Sea12b}. 

Therefore, since the average density profiles of constrained and unconstrained halos are slightly different, the same is must be true for their average velocity dispersion and anisotropy profiles. Specifically, the higher the background density, the higher the halo concentration, and hence, the steeper their density profile. Consequently, the steeper also the velocity dispersion and anisotropy profiles. 

Dividing the relations (\ref{sigma}) holding for constrained and unconstrained halos, we are led to 
\beq
\frac{\sigma\co(r)}{\sigma(r)}=\left[\frac{\rho\co(r)}{\rho(r)}\right]^{1/3}.
\eeq
In particular, at the virial radius where $\rho(r_0)=M_0 c^2/(4\pi r_0^3)/[f(c)(1+c)^2]$ we have
\beq
\frac{\sigma\co(r_0)}{\sigma(r_0)}=\left[\left(\frac{c\co}{c}\frac{1+c}{1+c\co}\right)^2\frac{f(c)}{f(c\co)}\right]^{1/3}.
\label{sigmna}
\eeq
Since the higher $\delta\m(t_0)$, the higher $c\co$, equation (\ref{sigma}) tells that the higher also $\sigma\co(r_0)$, in agreement with simulations. 


On the other hand, dividing the relations (\ref{beta0}) at the virial radius $r_0$ for the unconstrained and constrained halos, with concentrations $c$ and $c\co$, respectively, where the logarithmic slope of the density profile of NFW halos with $M_0$ and $r_0$ satisfies
\beq
\frac{\der\ln\rho}{\der \ln r}=-\left[1+\frac{2}{1+r_0/(rc)}\right], 
\label{dens}
\eeq
we arrive at 
\beq
\frac{\beta\co(r_0)}{\beta(r_0)}\approx \frac{0.1+1/[1+(c\co)^{-1}]}{0.1+1/(1+c^{-1})}.
\label{beta}
\eeq
Since the higher $\delta\m(t_0)$, the higher $c\co$, equation (\ref{beta}) tells that the smaller $\beta\co(r_0)$.

As can be seen in Figure \ref{all}, the biases in the velocity dispersion and anisotropy are the less marked among all the properties analyzed.

\subsection{Triaxial Shape}

As mentioned, the triaxial shape of halos, characterized by their ellipticity ($e\ge 0$) and prolateness ($|p|\le e$, with $p$ positive for oblate objects and negative for prolate ones), is related to that of the corresponding peaks through the kinematics of the final objects in a convoluted non-analytic way. However, \citet{Sea12b} showed that, globally, the shape of the isodensity contours in halos and protohalos vary with radius in a similar way. Moreover, since the deeper one goes in a halo, the less marked the influence of the kinematics in its shape, so the closer its elliptcity and prolateness to those of the corresponding protohalo. On the contrary, as one goes outwards, the more spherical haloes compared to their seeds. Thus, we will concentrate on the halo shape at small radii, for which simple analytic relations can be derived. 

The probability of a given ellipticity and prolateness of peaks is independent of their height $\nu$ and decreases with increasing curvature. In particular, the typical asphericity of peaks with $\delta_0$ at $R_0$ measured through the ellipticity $e$ diminishes with the average curvature $\lav x\rav$, according to (BBKS)
\beq
e\approx \frac{1}{\sqrt{5}}\left[\lav x\rav^2(R_0,\delta_0)+6/5\right]^{-1/2}.
\label{e}
\eeq

Since constrained peaks with $\delta_0$ at $R_0$ behave as unconstrained ones with $\delta_{\rm e0}=\delta_0-q\delta\m$ at $R_0$, their median ellipticity takes the form (\ref{e}) with the average curvature evaluated at $\delta_{\rm e0}$. Consequently, the ratio of median ellipticities {\it at small radii} (say, at $r\sim \rs$) of constrained and unconstrained halos with {\it large} $M_0$ at $t_0$, close to those of their corresponding peaks, is 
\beq
\frac{e\co}{e}\approx \left[\frac{\lav x\rav^2(R_0,\delta_0)+6/5}{\lav x\rav^2(R_0,\delta_{\rm e0})+6/5}\right]^{1/2}.
\label{ell}
\eeq 
Since $\lav x\rav(R,\delta)$ increases with increasing $\delta$ at fixed $R$, equation (\ref{ell}) implies that the larger $\delta\m(t_0)$, the larger the typical ellipticity $e\co$. In lower mass halos, the dependence of $e\co/e$ on $c\co/c$ is less simple, but the trend is similar. 

As shown in Figure \ref{all}, the ellipticity bias (at small radii and for large mass halos) is quite weak, which agrees with the results of simulations \citep{Cea20}. In particular, the reversal of the trend in the shape bias is found in simulations to take place at the same mass marking the frontier between the two similar regimes in the concentration bias \citep{FW10}, as predicted here.

The same derivation applied to the halo prolateness $p$ leads to a ratio of prolatenesses $p\co/p$ in constrained and unconstrained halos of the same form as $e\co/e$, but with the right hand member of equation (\ref{ell}) to the fourth power. Thus, the prolateness bias is predicted to be much more marked than the ellipticity bias.


\subsection{Spin}

The spin parameter $\lambda$ used in most studies of the secondary bias (e.g. \citealt{GW07}) is defined as \citep{Bea01}
\beq
\lambda=\frac{J}{\sqrt{2}M_0 r_0 V_0},
\label{lambdap}
\eeq 
where $J$ is the modulus (vector norm) of the total angular momentum (AM) ${\bf J}$ relative to the center of mass (c.o.m.) of the halo with $M_0$, and $V_0$ is its circular velocity at the radius $r_0$. According to the Tidal Torque Theory, in linear and moderately non-linear regime ${\bf J}$ is kept with the same direction and $J$ increases with time $t$ as $J(t)= a^2(t)\dot{D(t)}\,J^{\rm L}$, being  the $i$th component of the constant Lagrangian protohalo AM given by
\beq
J_{i}^{\rm L} = \epsilon_{ijk}{\bf T}_{jl}{\bf I}_{lk}
\label{1}
\eeq
where ${\bf T}$ is the Hessian of the potential at the c.o.m. of the protohalo at $\ti$ and {\bf I} is its inertia tensor \citep{W84}. Consequently, the ratio of the median AM of constrained and unconstrained halos with $M_0$ collapsed at the same time $t_0$, $J\co/J$, is simply equal to the ratio of their median Lagrangian AM modulus, $(J^{\rm L})\co/J^{\rm L}$. 

Using CUSP \citet{SM24} have recently derived the median Lagrangian protohalo AM for unconstrained halos with {\it virial} mass $M_0$ at $t_0$ in the peak model. The result is 
\beq
J^{\rm L}(M_0,t_0) \approx \frac{0.521}{\frac{5}{3}m+3} G\bar\rho^{1/3}_0 s^{\frac{5}{3}m}g M^{5/3}\frac{\delta}{D(\ti)},
\eeq
where $\bar\rho_0$ is the present mean cosmic density, $m=-(n+3)/2$, being $n\approx -1.75$ the effective power index of the CDM spectrum at galactic halo masses,
\beq
s^{-3}=\!\frac{2}{3 \pi}\!\left(\!\frac{n+5}{6}\!\right)^{3/2} G_0(\gamma,\gamma\nu)\,{\rm e}^{-\frac{\nu^2}{2}},
\label{im2}
\eeq
and
\beq
g=\left[\frac{(\gamma\nu)^{1/3}}{\lav x\rav^2(R_0,\delta_0)+6/5}\right]^2.
\eeq 
Note that $J^{\rm L}(M_0,t_0)$ is independent, indeed, of the (arbitrary) initial time $\ti$ because $\delta_0/D(\ti)$ is a function of $t_0$ alone (see eq.~[\ref{deltat}]). Therefore, taking into account that $V_0$ is the same in constrained and unconstrained halos with $M_0$ and $r_0$ (eq.~[\ref{v}]), we are led to 
\beq
\frac{\lambda\co}{\lambda}\!=\!\frac{J\co}{J}\!\approx\! \left[\frac{\lav x\rav^2(R_0,\delta_0)+6/5}{\lav x\rav^2(R_0,\delta_{\rm e0})+6/5}\right]^{2}.
\label{lambda}
\eeq
Since the higher $\delta\m$, the smaller $\lav x\rav(R_0,\delta_{\rm e0})$, and the larger the spin as found in simulations \citep{GW07}. In Figure \ref{all} we see that the spin is among the properties (together with concentration, formation redshift, and subhalo abundance) that show the most marked bias, in agreement with the results of simulations \citep{Mea18}. In addition, the common origin (directly related to the mean protohalo curvature) of the triaxial shape and spin biases is consistent with their observed correlation \citep{SP19}. 


\section{SUMMARY AND CONCLUSIONS}\label{dis}

Cosmological simulations show that halos with the same mass but different internal properties (concentration, formation time, velocity peak, subhalo abundance, kinematics, triaxial shape, and spin) are differently clustered, what is known as secondary  bias. 

Using the CUSP formalism relating halos with mass $M_0$ at the cosmic time $t_0$ with peaks with density contrast $\delta_0$ at scale $R_0$ in the Gaussian-smoothed random Gaussian density field at an initial (arbitrary) time $\ti$, we have examined the appealing idea suggested by \citet{Dea08} that the secondary bias could arise from the different typical curvature of peaks lying in different backgrounds have different typical curvatures. To do that, we have taken advantage that mergers can be ignored when dealing with the internal properties of halos, and focused on purely accreting objects. 

We have shown that the mean curvature of peaks with given $\delta_0$ and $R_0$ depends on their background density. In addition, we have demonstrated that the halo shape and spin directly arise from the curvature of the associated peaks, while all the remaining properties entering secondary bias arise from the curvature of peaks along the continuous $\delta(R)$ trajectory tracing the growth of accreting halos, which determine their density profile \citep{SM19}. Consequently, the halo shape, spin, and any property related to the density profile depend on the peak background, or equivalently, on the halo background. And, given the primary bias relating the local halo and matter densities (Paper I), this causes halos with different values of these properties to be differently clustered. 

We have found that the mean curvature of peaks constrained to lie in a background is essentially the same as for unconstrained peaks with a slightly different density contrast. This has allowed us to derive simple analytic expressions for the bias in all those halo properties. The predicted median values of these properties in halos lying in specific background densities, or equivalently, the clustering level of halos with specific values of the properties have been shown to agree with the trends found in simulations. 

Interestingly, the only difference between distinct properties is in their specific monotonic relation with the peak curvature. The peak curvature vs. background density, or equivalently, the clustering level of halos whose peaks have a specific curvature is obviously the same for all properties. Consequently, the reversal of the clustering level of halos as a function of the values of any given property from low to high halo masses found in simulations is a general feature. It is inherent to the dependence of curvature on the density contrast and scale of peaks, which does not depend on the particular property considered. Therefore, contrarily to what is commonly believed, such a reversal does not reflect any change with mass of the clustering trend of halos with different values of the property. 

Thus, the main conclusion of this work is that the secondary bias, like the primary one, is innate, and it is well reproduced in the peak model of structure formation.  

\begin{acknowledgments}
This work was funded by the Spanish MCIN/AEI/ 10.13039/501100011033 through grants CEX2019-000918-M (Unidad de Excelencia `Mar\'ia de Maeztu', ICCUB) and PID2022-140871NB-C22 (co-funded by FEDER funds) and by the Catalan DEC through the grant 2021SGR00679.
\end{acknowledgments}

{}

\end{document}